\documentclass[twocolumn,twocolappendix]{aastex631}
\usepackage{amsmath}
\usepackage{verbatim}
\usepackage{units}

\usepackage{subfigure}
\usepackage{units}
\usepackage{color}

\begin{document}

% \title{Unimolecular Chemical Kinetics in the Interstellar: Understanding the Dominance of Infrared Radiation over Pressure}
\title{ Unimolecular Chemical Kinetics in the Interstellar: Competition of Infrared Radiation and Collision Activation Mechanisms}

\author[0000-0001-8608-8485]{Xiaorui Zhao}
\affiliation{Center for Combustion Energy, Tsinghua University, Beijing 100084, P. R. China}
\affiliation{School of Aerospace Engineering, Tsinghua University, Beijing 100084, P. R. China}

\author[0000-0002-4880-6391]{Rui Ming Zhang}
\affiliation{Center for Combustion Energy, Tsinghua University, Beijing 100084, P. R. China}
\affiliation{Department of Energy and Power Engineering, Tsinghua University, Beijing 100084, P. R. China}

\author[0000-0002-2009-0483]{Xuefei Xu}
\affiliation{Center for Combustion Energy, Tsinghua University, Beijing 100084, P. R. China}
\affiliation{Department of Energy and Power Engineering, Tsinghua University, Beijing 100084, P. R. China}

\author[0000-0002-2863-7658]{Haitao Xu}
\affiliation{Center for Combustion Energy, Tsinghua University, Beijing 100084, P. R. China}
\affiliation{School of Aerospace Engineering, Tsinghua University, Beijing 100084, P. R. China}



\begin{abstract}

Unimolecular gas phase chemical reactions could be activated by both infrared (IR) radiation and inter-molecular collision in the interstellar environment. Understanding the interplay and competition between the radiation and collision activation mechanisms is crucial for assessing accurate reaction rate constants with an appropriate model. 
In this work, guided by an extended version of Lindemann theory that considers the contribution of both the radiation-activation and collision-activation to the rate constant of unimolecular reactions, we show that the relative importance of the two mechanisms can be measured by a dimensionless number $PR$ that is the ratio of the collision frequency to the radiation absorption rate of the molecule. The reaction kinetic is dominated by collision-activation or by radiation activation depending on whether $PR$ is larger or smaller than a reference value $PR^*$, which is determined to be $PR^* \approx 10$ based on magnitudes of molecular properties and is verified by detailed calculations of a number of typical interstellar unimolecular reactions. This method of evaluating the relative importance of the two mechanisms is checked against master equation calculations of two interstellar reactions: the dissociation reaction of silicilic acid around the asymptotic giant branch (AGB) star and the methyl association in Titan's atmosphere, and the validity is verified. 
The method can be used in the future to help determine the appropriate and effective modeling approach for chemical reactions in astrophysical environments.

\end{abstract}



\section{Introduction} \label{sec:intro}

Unimolecular reactions, typically dissociation and isomerization, as important elementary steps in many reaction networks, are ubiquitous in astrophysical environments. 
For unimolecular reactions to occur, the molecule must first reach an activated state by gaining internal energy. Two activation mechanisms have been proposed. The radiation-activation mechanism \citep{Perrin1919,Perrin1922} hypothesizes that the molecules are activated by absorbing energy from the temperature-dependent background blackbody infrared (IR) radiation, while the collision-activation mechanism \citep{Lindemann1922} suggests that the activation energy comes from collisions with surrounding molecules.

Historically, although the radiation-activation mechanism was proposed earlier, it was challenged by the facts that the blackbody radiation intensity of the environment is too low to trigger any reactions in practice and it cannot explain the observed pressure dependence of unimolecular reaction rates \citep{Langmuir1920}. 
In contrast, the collision-activation mechanism led to the Lindedmann theory \citep{Lindemann1922} that gave quantitative prediction of the pressure dependence of unimolecular reaction rates by introducing a collision-induced energetic complex between reactant and product, 
%In 1922, the Lindedmann theory \citep{Lindemann1922} was established which follows a collision activation mechanism and gives a good explanation to the pressure dependence of unimolecular reaction rates by introducing a collision-induced energetic complex between reactant and product, 
providing a basic framework for subsequent refinement theories widely used today for unimolecular reaction kinetics, like the Rice-Ramsperger-Kassel-Marcus (RRKM) theory \citep{Marcus1952} and the more sophisticated master equation method \citep{Klippenstein2002,Pilling2003,Glowacki2012,Georgievskii2013,Zhang2022}.

In 1990s, ion dissociation activated thermally through radiation absorption under extreme low-pressures were reported with pressure-independent reaction rates, which provided the first clear evidences to support the infrared-radiation activation theory and reestablished radiation activation \citep{Kofel1988,Dunbar1995,Dunbar1998,Price1996,Klippenstein1996} as a competing mechanism for unimolecular reactions.
%Hereafter, collision and radiation mechanisms have been recognized as two competing mechanisms for activating the chemical reactions. 
While collision activation is the dominant mechanism under regular conditions, like in atmosphere chemistry and combustion chemistry, infrared-radiation activation may play an important role in some special cases, for example,  the unimolecular reactions occurring at extremely low pressures.

\begin{figure}[htbp]\label{Figure:Fig0}
\includegraphics[width=0.9\columnwidth]{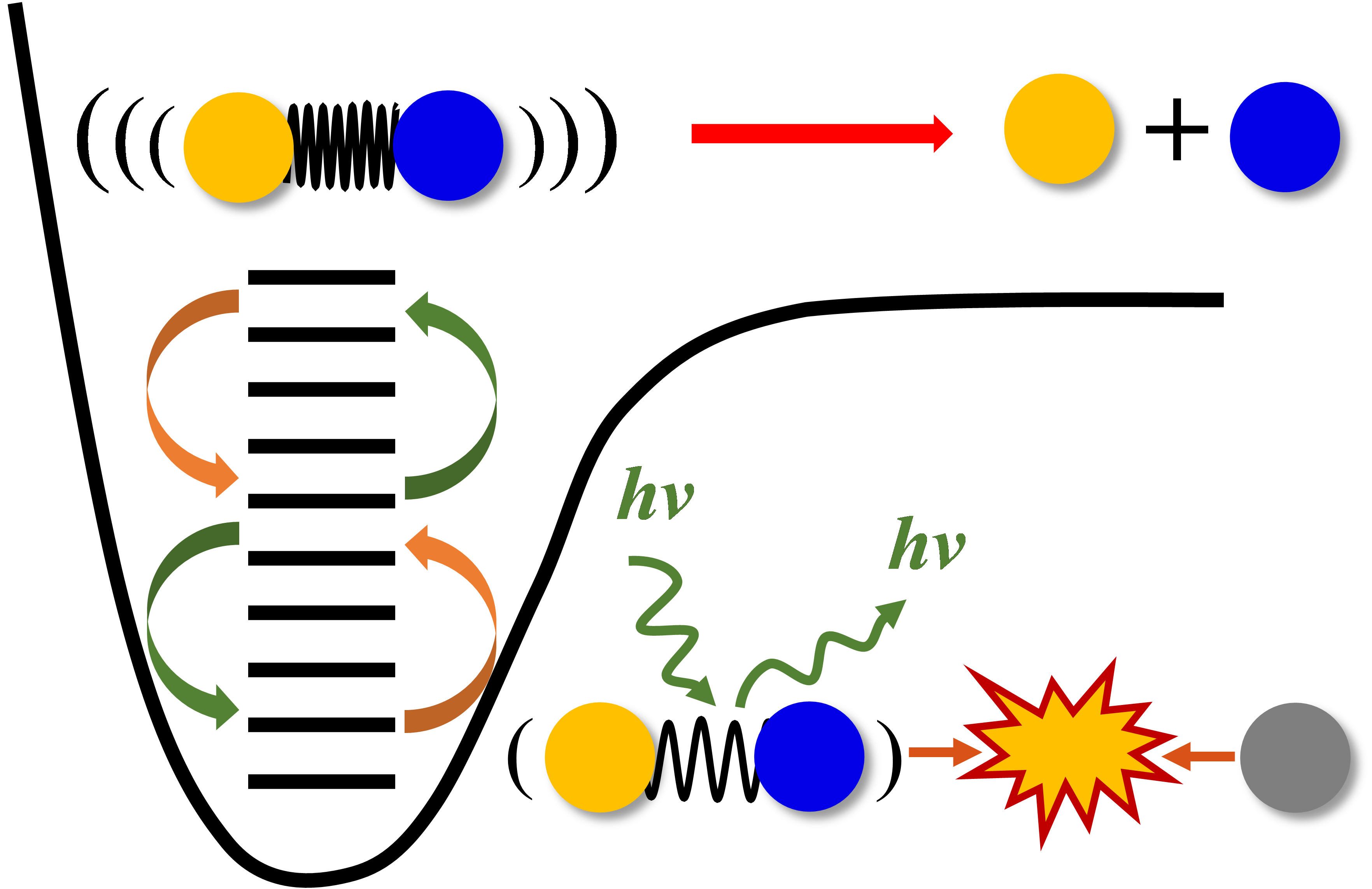}
\centering
\caption{Schematic diagram of the unimolecular reaction activation mechanism. Orange, green, and red arrows denote collision and radiation induced activation and deactivation processes, and chemical reaction process, respectively.} 
\end{figure}

In interstellar environments, gas-phase reactions often occur over a wide range of pressures, from dense planetary atmospheres to highly rarefied interstellar media. %therefore both these two activation mechanisms may determine the kinetics. 
It has been reported that in certain interstellar environments, gas-phase reaction rates and species abundance may exhibit notable discrepancies from theoretical predictions that consider only the collision activation. \cite{Vuitton2012} and \cite{Plane2022} demonstrated that when considering additionally the role of background IR radiation activation for unimolecular reactions, the theoretically calculated rate constants of the gas-phase reactions occurring in planetary atmospheres and in stellar outflows at low pressures would increase by several orders of magnitude, highlighting the potential significance of background IR radiation in influencing interstellar reaction kinetics.

The collision-activation and radiation-activation mechanisms can be taken into account either separately or combined in the Master equation (ME) modeling to predict temperature- and pressure-dependent reaction rate constants. To efficiently calculate rate constants, it is of importance to better understand the competition between the two activation mechanisms and the dominance of one over the other in order to adopt the appropriate mechanism in a specific condition.

In this work, we first use the unimolecular dissociation reaction of silicilic acid:
\[
{\rm OSi(OH)}_2 \rightarrow {\rm SiO}_{2} + {\rm H}_2{\rm O}     \qquad\qquad     ({\rm R}1) 
\]
as a model system to present the general behavior of unimolecular reaction kinetics under the combined effects of two activation mechanisms over a wide range of pressures, with a focus on the understanding the competition between the two mechanisms. 
Based on the results, we propose a simple approach to distinguish collision-dominated and radiation-dominated regimes, which can be used to facilitate the modeling of interstellar chemical reactions.
To demonstrate the feasibility of the proposed approach, we apply it to kinetics study of two reactions in two real interstellar environments: R1 in the asymptotic giant branch (AGB) star surroundings and methyl association reaction:
\[ 
{\rm CH}_{3} + {\rm CH}_{3} \rightarrow {\rm C}_{2}{\rm H}_{6}           \qquad\qquad        ({\rm R}2) 
\]
in the Titan's atmosphere. 
For these two cases, the regions where one mechanism dominates are determined using our approach, and then the reaction constants in the corresponding regions are calculated using the ME with only the dominating mechanism, and the results are in good agreement with the reaction constants calculated using the ME with both collision and radiation mechanisms considered.

\section{Methodology} \label{sec:theory}

In this section, we first describe how to obtain the rate constants of unimolecular dissociation reactions controlled by both collision (pressure) and radiation from the master equation (ME) method. We note that although we focus on the rate constants of the unimolecular reactions in this work, the ME method is versatile, i.e., applicable to various types of reaction systems, not just unimolecular reactions, and comprehensive, i.e.,  capable of providing detailed information of the reaction mechanism and kinetics. The ME results with appropriate treatments are usually used as the benchmark in related studies.
On the other hand, the ME method is computationally expensive and time-consuming.
For unimolecular reactions, an extended version of Lindeman theory (ELT) \citep{Price1996} provides a convenient analytical framework to consider dependence of the rate constants on both the pressure and IR radiation. We thus also give an introduction to the ELT theory and how to determine the four unknown quantities required by the theory.

\subsection{Master equation method}

We take the unimolecular dissociation reaction as an example to explain the ME method, and more details can be found in refs \citep{Georgievskii2013,Zhang2022}. 
The internal energy of the unimolecular reactant can be discretized into finite-width energy bins (also called energy grains) to represent various microscopic energy states.
The master equation is a set of equations that describes the evolution of the populations of all reactants and products in all the energy bins, from which macroscopic kinetics properties of the reaction system can be obtained.
In this work, we neglect the recombination of the bimolecular product into the reactant, which holds 
% if the concentration of the unimolecular reactant is very small compared to other species (the bath gases) or 
if the time scale of the concentration variation of the bimolecular species is much slower than that of energy relaxation \citep{Georgievskii2013,Zhang2020}, and such situation is relevant to rarefied astrophysical environment.
In such case, the ME method only needs to consider the population of the unimolecular reactant in its internal energy bins.
When both collision and radiation induced energy transfers are considered (In this work we use ``radiation'' to represent the background blackbody IR radiation and not touching photochemistry, which refers to the excitation of the reactants to higher electronic states by more energetic radiation), the time evolution of $\rho_i$, the population of the unimolecular reactant in an energy bin $i$, can be expressed as
\begin{equation}\label{eq:ME}
    \begin{split}
        \frac{d\rho_{i}}{dt} = Z\sum_{j}^{}{\lbrack\ P(j \rightarrow i)\rho_{j} - P(i \rightarrow j)\rho_{i}\ \rbrack} \\
        + \sum_{j}^{}{\lbrack\ R(j \rightarrow i)\rho_{j} - R(i \rightarrow j)\rho_{i}\ \rbrack} - k_{i}\rho_{i}
    \end{split}
\end{equation}
in which $Z$ is the collision frequency that can be computed with hard-sphere collision model or Lennard-Jones collision model \citep{Glowacki2012,Georgievskii2013,Zhang2022}; 
$P(j \rightarrow i)$ is the collisional energy transfer probability from bin $j$ to bin $i$, which is usually calculated with the exponential down model \citep{Glowacki2012,Georgievskii2013,Zhang2022}; $R(j \rightarrow i)$ is the radiation induced transition rate coefficient from bin $j$ to bin $i$, which can be computed from Einstein coefficients \citep{Plane2022,Salzburger2022} describing the absorption or emission probability of photons by molecules (details can be found in Appendix A); and $k_{i}$ is the microcanonical dissociation rate coefficient of the reactant at energy bin $i$, which can be computed using the RRKM theory. 
The first two terms on the right hand side of Eq.~\eqref{eq:ME} denote the rates of energy state transition due to collision with the bath gases and due to radiation absorbtion, respectively, and the third term is the rate of depletion due to dissociation.

The master equation of the unimolecular reaction can be concisely written in matrix form as:
\begin{equation}\label{eq:ME_matrixform}
    \frac{d\boldsymbol{\rho}}{dt} = - \boldsymbol{{\rm W }\rho}
\end{equation}
where \(\boldsymbol{\rho}\) is the energy bin population vector, and \(\boldsymbol{\rm W }\) is the transition matrix accounting for the contributions from the collisional transition, the radiation-induced transition, and the reaction, with elements: 
\begin{equation}\label{eq:ME_W}
    \begin{split}
             W _{ij} = - ZP(j \rightarrow i) + Z\sum_{i'}^{}{P(i \rightarrow i')}\delta_{ij} \\ 
            - R(j \rightarrow i) + \sum_{i'}^{}{R(i \rightarrow i')}\delta_{ij} + k_{i}\delta_{ij}
    \end{split}
\end{equation}
Then by solving the master equation via diagonalization of the transition matrix, the temporal evolution of the population can be yielded \citep{Glowacki2012,Georgievskii2013,Zhang2022}. 
The rate constant can be extracted with chemically significant eigenmodes (CSE) theory \citep{Widom1965,Bartis1974}, which separates eigenvalues of the transition matrix into chemically significant eigenmodes and internal-energy-relaxation eigenmodes according to their magnitudes.
For this single-well single-channel dissociation system, the smallest eigenvalue directly corresponds to the dissociation rate constant.

\subsection{Extended Lindemann theory}

In the classical Lindemann theory, the dissociation of the reactant A to products P involves three processes: 
\begin{equation}\label{eq:Process_M1}
    {\rm A}\  + \ {\rm M}\ \stackrel{k^{\rm c}_{1}}{\longrightarrow}\ {\rm A^{*}}\  + \ {\rm M}
\end{equation}
\begin{equation}\label{eq:Process_M-1}
    {\rm A^{*}}\  + \ {\rm M}\ \stackrel{k^{\rm c}_{-1}}{\longrightarrow}\ {\rm A}\  + \ {\rm M}
\end{equation}
\begin{equation}\label{eq:Process_k}
    {\rm A^{*}}\ \stackrel{k_{2}}{\longrightarrow}\ {\rm P}
\end{equation}
which represent the activation of A to its energetic state A* by collisions with a molecule M, the de-activation of A* back to A via collisions, and the decomposing of A* into P, respectively. The molecule M could be any species in the system, including the bath gas molecules or the reactant A itself.

Typically, the concentration of the activated molecules A* changes over a much longer time scale than other species and thus can be considered as at a quasi-steady state, 
i.e., $d \lbrack {\rm A^*} \rbrack /dt \approx 0$. Under this approximation, the rate constant for the unimolecular dissociation of A can be expressed as:
\begin{equation}\label{eq:LindemannP}
    k_{\rm{uni}} = k_{2}\frac{k^{\rm c}_{1}\lbrack {\rm M} \rbrack}{k^{\rm c}_{- 1}\lbrack {\rm M} \rbrack + k_{2}},
\end{equation}
in which $\lbrack {\rm M} \rbrack$ is the concentration of M, $k^{\rm c}_{1}$ and $k^{\rm c}_{-1}$ are the rate coefficients of the collisional activation and de-activation processes, and $k_{2}$ is the rate coefficient of the decomposition of A* to products.

The unimolecular reaction rate constant given by Eq.~\eqref{eq:LindemannP} is in general pressure-dependent because the bath gas concentration [M] is proportional to the pressure $p$ of the system. There are two regimes where the dependence is straightforward. When the pressure is very high, such that $\lbrack {\rm M} \rbrack \gg  \frac{k_{2}}{k^{\rm c}_{- 1}} $, then the unimolecular dissociation is controlled by the competition between collisional activation and collisional de-activation. At this high-pressure limit, the reaction rate constant $k_{\rm{uni}}$ is pressure-independent, with the value ${k}_{\rm uni}|_{p{\rightarrow}\infty} = {k}^{\rm c}_{\infty} = \frac{k_{2} k^{\rm c}_{1}}{k^{\rm c}_{- 1}}$. 
On the other hand, in the low pressure range, $\lbrack {\rm M} \rbrack \ll \frac{k_{2}}{k^{\rm c}_{- 1}} $, thus ${k}_{\rm uni}|_{p{\rightarrow}0} = {k}^{\rm c}_{0} = k^{\rm c}_{1}\lbrack {\rm M} \rbrack$, which means that the rate constant varies linearly with [M] or pressure \textit{p} and the dissociation is dominated by the collisional activation step.

% As shown in Figure 1, besides gaining energy to the activated state via collisions with the bath gas molecules, the molecule A can also be activated by absorbing background IR radiation. 
The extended Lindemann theory includes the radiation-activation mechanism in the framework of Lindemann theory, that is, together with Eqs.~\eqref{eq:Process_M1}, \eqref{eq:Process_M-1}, and \eqref{eq:Process_k}, the radiational activation and de-activation processes are considered \citep{Price1996}:
\begin{equation}\label{eq:Process_hv1}
    {\rm A}  + h\nu\ \stackrel{k^{\rm r}_{1}}{\longrightarrow}\  {\rm A^{*}}
\end{equation}
\begin{equation}\label{eq:Process_hv-1}
     {\rm A^{*}}\ \stackrel{k^{\rm r}_{- 1}}{\longrightarrow}\ {\rm A}  + h\nu
\end{equation}
in which $k^{\rm r}_{1}$ and $k^{\rm r}_{- 1}$ are the radiation-activation and de-activation rate coefficients, respectively.

Equations \eqref{eq:Process_M1}-\eqref{eq:Process_hv-1} form the reaction system of the ELT theory. Before dive into the full system, it is illuminating to examine the properties of dissociation when only the radiation-activation mechanism is considered, i.e., the system is consisted of only Eqs.~\eqref{eq:Process_k}, \eqref{eq:Process_hv1},and \eqref{eq:Process_hv-1}. It is easy to see that 
%under the steady-state approximation, 
the unimolecular dissociation reaction rate constant of this system is
\begin{equation}\label{eq:Lindemannhv}
    k_{\rm{uni}} = k_{2}\frac{k^{\rm r}_{1}\lbrack h\nu \rbrack} 
    { k^{\rm r}_{- 1}\lbrack h\nu \rbrack + k_{2}}
\end{equation}
where $\lbrack h\nu \rbrack$ represents the concentration of photons that participate in the radiational activation/de-activation process.

Comparing Eq.~\eqref{eq:Lindemannhv} and Eq.~\eqref{eq:LindemannP}, we see the remarkable resemblance between $\lbrack h\nu \rbrack$ and $\lbrack {\rm M} \rbrack$ in determining the behavior of the rate constant $k_{\rm{uni}}$.
When the active-photon-concentration $\lbrack h\nu \rbrack$ is very high, ${k}_{\rm uni}|_{[h\nu]{\rightarrow}\infty} = {k}^{\rm r}_{\infty} = \frac{k_{2} k^{\rm r}_{1}}{k^{\rm r}_{- 1}}$, which is independent of $\lbrack h\nu \rbrack$; while when $\lbrack h\nu \rbrack$ is very low, ${k}_{\rm uni}|_{[h\nu]{\rightarrow}0} = {k}^{\rm r}_{0} = k^{\rm r}_{1}\lbrack h\nu \rbrack$, which changes linearly with $\lbrack h\nu \rbrack$.
The difference between the collision-activation case and the radiation-activation case is that the concentration of the bath gas molecules $\lbrack {\rm M} \rbrack$ depends on pressure $p$, but the concentration of the active-photon $\lbrack h\nu \rbrack$ depends on the background IR radiation strength that varies with the temperature of the environment and the property of the molecule A.
For example, \cite{Dunbar1994} and \cite{Dunbar1998} proposed that for large molecules with abundant IR active vibrational modes, more photons could be absorbed/emitted, thereby participating in the reaction activation, leading to a very high $\lbrack h\nu \rbrack$ and consequently constant ${k}_{\rm uni}$. They named this high active-photon-concentration limit as the ``large-molecule limit'' and the corresponding low active-photon-concentration limit for molecules with few IR active vibrational modes as the ``small-molecule situation'' \citep{Dunbar1994,Dunbar1998}.

When both collision-activation and radiation-activation are considered (referred to as the combined mechanism in the rest of the paper), the ELT method gives the unimolecular reaction rate constant as a function of both $\lbrack {\rm M} \rbrack$ and $\lbrack h\nu \rbrack$ \citep{Price1996}
\begin{equation}\label{eq:LindemannELT}
    k_{\rm{uni}} 
    = k_{2}\frac{k^{\rm c}_{1}\lbrack {\rm M} \rbrack + k^{\rm r}_{1}\lbrack h\nu \rbrack}{k^{\rm c}_{- 1}\lbrack {\rm M} \rbrack + k^{\rm r}_{- 1}\lbrack h \nu \rbrack + k_{2}} 
    = \frac{k^{\rm c}_{1}\lbrack {\rm M} \rbrack + k^{\rm r}_{1}\lbrack h\nu \rbrack}
    {\frac{k^{\rm c}_{1}\lbrack {\rm M} \rbrack}{{{k}}^{\rm c}_{\infty}} + \frac{k^{\rm r}_{1}\lbrack h\nu \rbrack}{{{k}^{\rm r}_{\infty}}} + 1} 
\end{equation}
This result shows the competition between pressure and radiation on determining the unimolecular chemical kinetics, which will be discussed in detail later. Here we note that the reaction rate constant given by the ELT theory is completely specified by four quantities: ${k}^{\rm c}_{\infty}$, ${k}^{\rm r}_{\infty}$, $k^{\rm c}_{1}\lbrack {\rm M} \rbrack$, and $k^{\rm r}_{1}\lbrack h\nu \rbrack$, 
which can fitted in the Arrhenius form from the experiment data or theoretical results \citep{Vuitton2012}.  
In this work, we determine these quantities in a computationally efficient approach by adopting approximate analytical forms of the internal energy distribution of the unimolecular reactant under the reaction condition.

We first note that at the high-pressure limit or the ``large-molecule limit'', the internal energy is in Boltzmann distribution, because the thermal equilibrium distribution of the reactant is maintained because the relatively slow dissociation compared to the high collision frequency or the strong radiation absorption. Thus we have
\begin{equation}\label{eq:k_infty}
    k^{\rm c}_{\infty} = k^{\rm r}_{\infty} = \sum_{E_{i}}^{}{N_{\rm{BD}}(E_{i}) \cdot k_{i}}
\end{equation}
where $N_{\rm{BD}}(E_{i})$ is the Boltzmann probability of the energy bin $E_{i}$ and $k_{i}$ is the corresponding microcanonical reaction rate coefficient of the reactant.
Away from that limit, dissociation decreases the populations in the high-energy tail, especially with energy levels higher than the activation barrier. 
At the low-pressure limit or the ``small-molecule limit'', the dissociation process is assumed to be fast relative to the slow internal energy relaxation processes once the the internal energy of species is above the barrier, i.e.,  as long as a molecule is pumped to an energy state above the dissociation barrier, it dissociates immediately, which means that the populations of the high-energy states are essentially zero.
A well-accepted approximation of the energy distribution in those cases is the ``truncated Boltzmann distribution (TBD)'' \citep{Dunbar1991,Dunbar1994}, which, as the name suggests, sets the populations of the energy levels below the barrier to be the Boltzmann distribution while above the barrier to be zero, i.e.
\begin{equation}
N_{\rm{TBD}}(E_{i}) = \left\{ 
\begin{aligned}
& \frac{N_{\rm{BD}}(E_{i})}{\sum_{E_{i} < E_{0}}^{}{N_{\rm{BD}}(E_{i})}} & E_{i} < E_{0}  \\
& 0 & E_{i} > E_{0} 
\end{aligned} 
\right. 
\end{equation}
where $E_{0}$ is the energy barrier.
%In section 4.1, we will validate the TBD approximation.}

Based on the TBD approximation, the unimolecular dissociation rate in the low-pressure limit or the small-molecule limit is given by the rate of ``pumping'' the internal energy above the reaction barrier \citep{Dunbar1994}, i.e.,
\begin{equation}
    k_{\rm{uni}} \cong k_{\rm{pump}} = \sum_{E_{i} < E_{0}}^{}{N_{\rm{TBD}}(E_{i}) \cdot \phi_{i}}
\end{equation}
where $\phi_{i}$ is the rate of pumping the molecules at the energy level $E_{i}$ to a level above the barrier. 
For collisional activation, the pumping rate $\phi_{i}^{\rm{c}}$ can be estimated as
\begin{equation} \label{eq:Ri_collision}
    \phi_{i}^{\rm{c}} = Z \sum_{E_{j} > E_{0}}^{}{P(i \rightarrow j)} ,
\end{equation}
which gives
\begin{align}
   &  k^{\rm c}_{1}\lbrack {\rm M} \rbrack = k_{\rm{pump}}^{\rm c} \nonumber \\
    & = Z \sum_{E_{j} > E_{0}} {\sum_{E_{i} < E_{0}} \frac{N_{\rm{BD}}\left( E_{i} \right)  }{\sum_{E_{i} < E_{0}} {N_{\rm{BD}}\left( E_{i} \right) }} P(i \rightarrow j) }
\label{eq:kpump_collision}
\end{align}
%\begin{equation}
%\label{eq:kpump_collision}
%    k^{\rm c}_{1}\lbrack {\rm M} \rbrack\ = k_{\rm{pump}}^{\rm c} 
%    = Z \sum_{E_{j} > E_{0}} {\sum_{E_{i} < E_{0}} \frac{N_{\rm{BD}}\left( E_{i} \right)  }{\sum_{E_{i} < E_{0}} {N_{\rm{BD}}\left( E_{i} \right) }} P(i \rightarrow j) }
%\end{equation}
Similarly, for radiational activation, the pumping rate $\phi_{i}^{\rm{r}}$ is
\begin{equation}\label{eq:Ri_radiation}
    \phi_{i}^{\rm{r}} = \sum_{E_{j} > E_{0}}^{}{R(i \rightarrow j)}
\end{equation}
and
\begin{align}
    & k^{\rm r}_{1}\lbrack h\nu \rbrack = k_{\rm{pump}}^{\rm{r}}  \nonumber \\
    & = \sum_{E_{j} > E_{0}}^{}{\sum_{E_{i} < E_{0}}^{}\frac{N_{\rm{BD}}\left( E_{i} \right)}{\sum_{E_{i} < E_{0}}^{}{N_{\rm{BD}}\left( E_{i} \right)}}R(i \rightarrow j)}
\label{eq:kpump_radiation}
\end{align}

Using Eqs.~\eqref{eq:LindemannELT}, \eqref{eq:k_infty}, \eqref{eq:kpump_collision}, and \eqref{eq:kpump_radiation}, we can efficiently obtain the four quantities in the ELT with the combined mechanism.
% and evaluate the dominance of collision-induced and radiation-induced activation mechanisms under different reaction conditions. 

% The computational details are given in the supporting information. 

\section{computational details} \label{sec:computation}

In the ME and ELT calculations of reaction R1, we use the electronic structure information (including the optimized geometries, energies, harmonic vibrational frequencies, and the spontaneous Einstein coefficients of reactants and products) and the high-pressure-limit rate coefficients reported by Plane and Robertson \citep{Plane2022}, and we also use the same kinetics modelling settings as theirs. 
We extend the calculations to wider temperature and pressure range.

The Reaction R2 is the reverse reaction of ethane dissociation, and the pressure and temperature dependent reaction rate constant of R2 is determined by the reverse unimolecular dissociation reaction rate constant and the temperature-dependent equilibrium constant.
In the ME and ELT calculations of reaction R2, we use the electronic structure information (the optimized geometries, energies, and harmonic vibrational frequencies) and the collisional energy transfer model used in \citet{Klippenstein1999} and \citet{Klippenstein2006}. 
The spontaneous Einstein coefficients needed in the radiational energy transfer modelling are obtained from \citet{Vuitton2012}. 
The densities of states are calculated by using the rigid-rotor and harmonic-oscillator approximation, and the microcanonical rate coefficients of the unimolecular dissociation reaction are obtained by performing the inverse Laplace transformation to the high-pressure-limit reaction rate coefficients of methyl radicals association reaction in the form of \(k_{\rm{rec},\infty} = 5.26 \times 10^{- 10}\ {(T/30.2)}^{- 0.359}\ \rm{{cm}^{3}{molecule}^{- 1}s^{- 1}} \) \citep{Vuitton2012}.
%respectively, rather than the more complex hindered rotor treatment and variable reaction coordinate transition state theory (VRC-TST) in the references \citep{Klippenstein1999,Klippenstein2006}. 

%For R1 and R2, the ME calculation with equations \ref{eq:ME}, \ref{eq:ME_matrixform}, and \ref{eq:ME_W} as well as the ELT calculation with equations \ref{eq:LindemannELT}, \ref{eq:k_infty}, \ref{eq:kpump_collision}, and \ref{eq:kpump_radiation} use the same electronic structure calculation results and the kinetics modelling settings introduced above. 

In the test calculations for the reference value of dimensionless number identifying the turning point between the collision- and radiation-dominated regions, we simply use the  B3LYP \citep{Becke1993,Lee1988,Becke1988}/6-31g(d)\citep{631gd} method for calculating electronic structure information of the dissociation reactions of four hydrocarbons including the optimized structures, energies, and harmonic oscillator frequency and strength of vibrational modes. For the isomerization reactions of cyanomethanimine, the electronic structure information of are obtained from \citet{Vazart2015a}.
%Several typical unimolecular reactions have been tested in section 4.2. For these reactions, we only need to get the terms \(k^{\rm c}_{1}\lbrack {\rm M} \rbrack \) and \(k^{\rm r}_{1}\lbrack h\nu \rbrack\) to evaluate the competition between collisional and radiational activation, rather than do the total ME or ELT calculation. 
%For the isomerization reactions of two cyanomethanimine isomers, the electronic structure calculation results are taken from reference \citep{Vazart2015a}, and the kinetics modelling settings are consistent with R1. 
The calculations of the density of states,  the collision energy transfer model and parameters, and the radiational energy transfer model refer to R2 for the hydrocarbon reactions and to R1 for the cyanomethanimine isomerization reactions.

The electronic structure calculations are performed with the Gaussian software package \citep{g16}. All ME calculations are carried out with the open source code Tsinghua University Minnesota Master Equation program (TUMME) \citep{Zhang2022,TUMME2023}, of which the new features involving IR radiation activation mechanism will soon be publicly available in a future release.

\section{Result and Discussion} \label{sec:result}

\subsection{Collision-Radiation combined activation mechanism}

The silicilic acid dissociation reaction (R1) is common around AGB stars, contributing to the interstellar dust formation \citep{Plane2013,Plane2022}. 
In this subsection, we take it as a model system and study its kinetics at three temperatures (1000 K, 1500 K, and 2500 K) typical in the AGB star surrounding environment, over a wide range of pressure, from 10\textsuperscript{-10} to 10\textsuperscript{10} bar, to demonstrate the combined effect of pressure and radiation on the unimolecular kinetics and to validate the ELT theory.
% for validating the newly proposed approximate treatment in Lindermann theory framework.

%% The "ht!" tells LaTeX to put the figure "here" first, at the "top" next
%% and to override the normal way of calculating a float position
\begin{figure}[ht!]\label{Fig1}
\includegraphics[width=0.9\columnwidth]{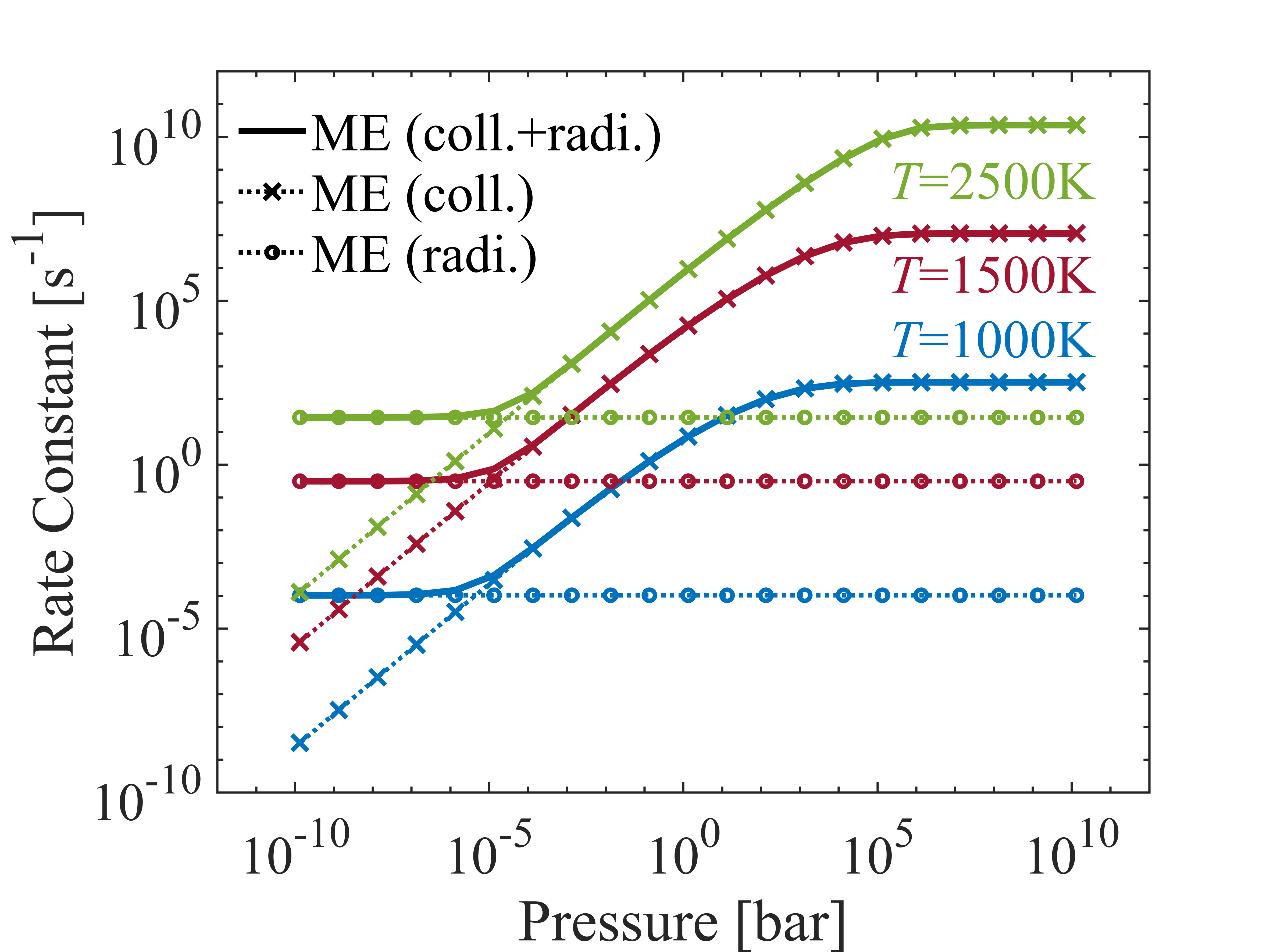}
\centering
\caption{Rate constants of the silicilic acid dissociation reaction as a function of pressure. %calculated by the master equation (ME) method. 
%The solid lines with dots represent the results taking both collision and radiation into account, 
%and the dashed lines with crosses denote the results only considering collision. 
}
\end{figure}

Figure 2 plots the R1 reaction rate constants obtained by the ME calculation Using the collision-radiation combined activation mechanism as a function of pressure, as well as those obtained by only considering the collision activation mechanism or the radiation activation mechanism. %for comparison.

From the comparison, %By comparing these results obtained using different mechanisms in Figure 2, 
we can clearly see that at a specific temperature, according to the environmental pressure, %values, 
the kinetics can be divided into the ``collision-dominant regime'', the ``radiation-dominant regime'', and the ``collision-radiation combined activation regime'' that lies between the other two. 

In the collision-dominant regime, the result of the collision activation mechanism coincides with that of the combined activation mechanism, that is, when the pressure is higher than a threshold, the rate constant doesn't vary with pressure, which is the high-pressure limit (around 10\textsuperscript{5} to 10\textsuperscript{6} bar for the current case, depending on the temperatures); and when the pressure is low enough, the rate constant falls off linearly with the decreasing pressure, meaning that the low-pressure limit of the collision mechanism is reached.

When the pressure is further decreased, in the radiation-dominated regime, the rate constant is pressure-independent again, instead of the linear dependence on pressure predicted by the collision activation mechanism, because the radiation activation surpasses the collision activation at these extremely low pressures. Moreover, the importance of including the radiation activation is highlighted in the radiation-dominated regime, in which the rate constant would be grossly underestimated if only taking the collision activation into account.
Figure 2 also suggests that although either collision activation or radiation activation mechanism can be solely used to predict the kinetics in the ``pressure-radiation combined action regime'' to an approximately satisfactory extent (underestimating the rate by about a factor of two), to get more accurate depiction, the combined mechanism is required.

Figure 3 compares the ME results using the combined mechanism with those obtained by the ELT method for reaction R1 at $T=\unit[1500]{K}$. 
%The black solid curve with circles is the solution of master equation, and the magenta dashed curve with triangles is the prediction with the extended Lindemann theory framework. 
We can see that the ELT result reproduces the general picture of collision-radiation competition kinetics revealed by the ME calculation, 
and it provides relatively quantitative prediction in the radiation-dominated regime and in the collision-radiation combined activation regime.

\begin{figure}[ht!]\label{Figure:Fig2}
\includegraphics[width=0.9\columnwidth]{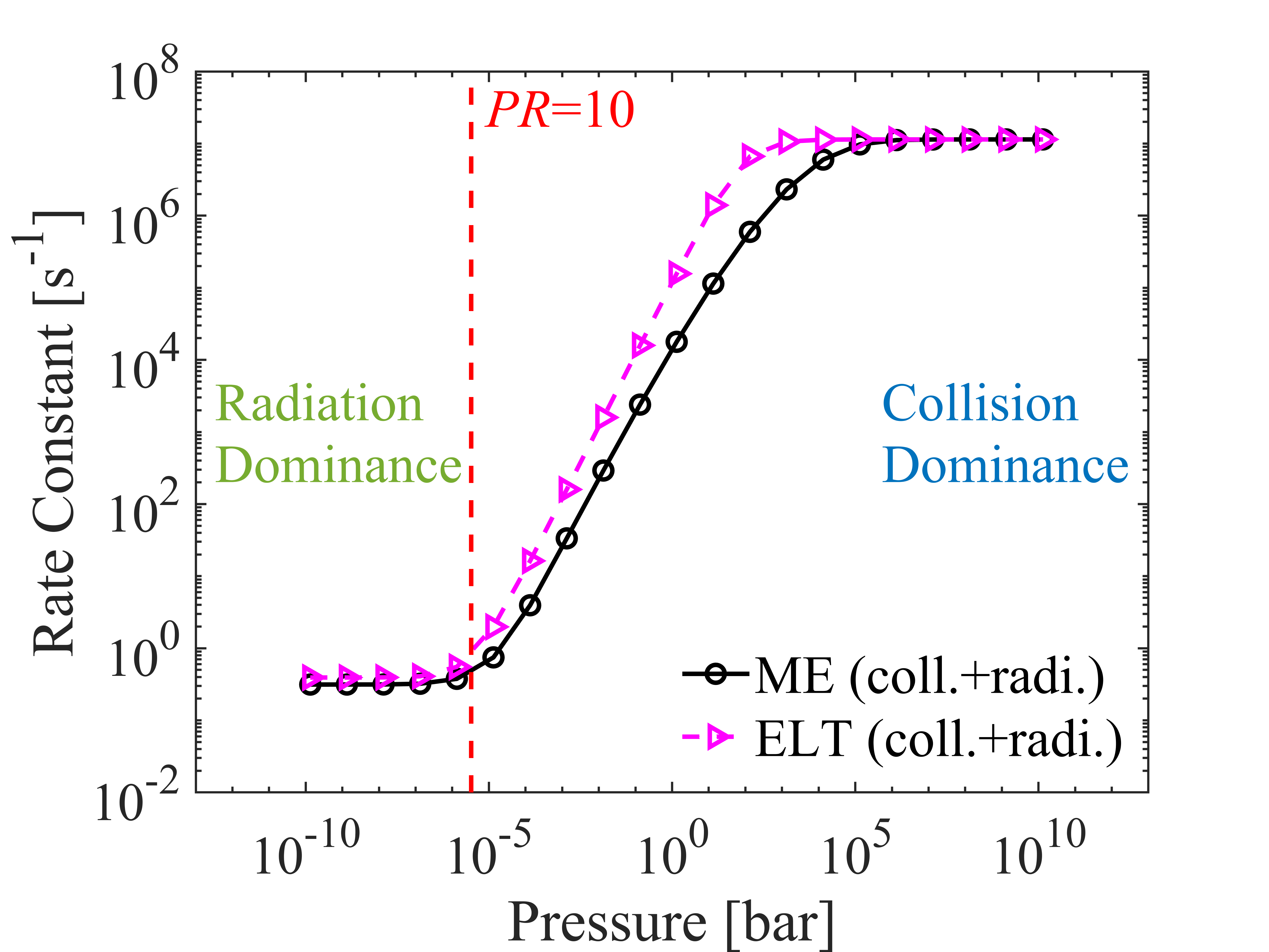}
\centering
\caption{Comparison of the master equation method using the combined mechanism and the ELT theory in predicting the rate constants of silicilic acid dissociation reaction as a function of pressure at $T=1500K$.
The red dashed line marks the pressure corresponding to $PR = 10$.}
\end{figure}

The success of the ELT method is partially attributed to the correct understanding of the internal energy distribution under the reaction condition, which is confirmed by plotting the internal energy distribution of silicilic acid at three representative pressures ($\unit[10^{-9}]{}$, $\unit[10^{-1}]{}$, and $\unit[10^{9}]{bar}$) and $\unit[1500]{K}$ calculated by the ME method. 
As shown in Figure 4, at $\unit[10^{-9}]{}$ and $\unit[10^{-1}]{bar}$, where the system is in both the ``small molecule limit" of radiation absorption and the low pressure limit, below the reaction barrier (64.96 kcal/mol in current reaction) the population distribution of the internal energy satisfies the thermal Boltzmann distribution, and above the barrier the distribution is truncated, which supports the TBD assumption. 
At $\unit[10^{9}]{bar}$, where the system reaches the high pressure limit, the thermal equilibrium of silicilic acid is approached due to the high collision frequency, and the Boltzmann distribution is obtained for all the energy bins.
%This proves our correct understanding of the internal energy distribution picture, which is significant for the parameters calculation in the extended Lindemann theory.

\begin{figure}[ht!]\label{Figure:Fig3}
\includegraphics[width=0.9\columnwidth]{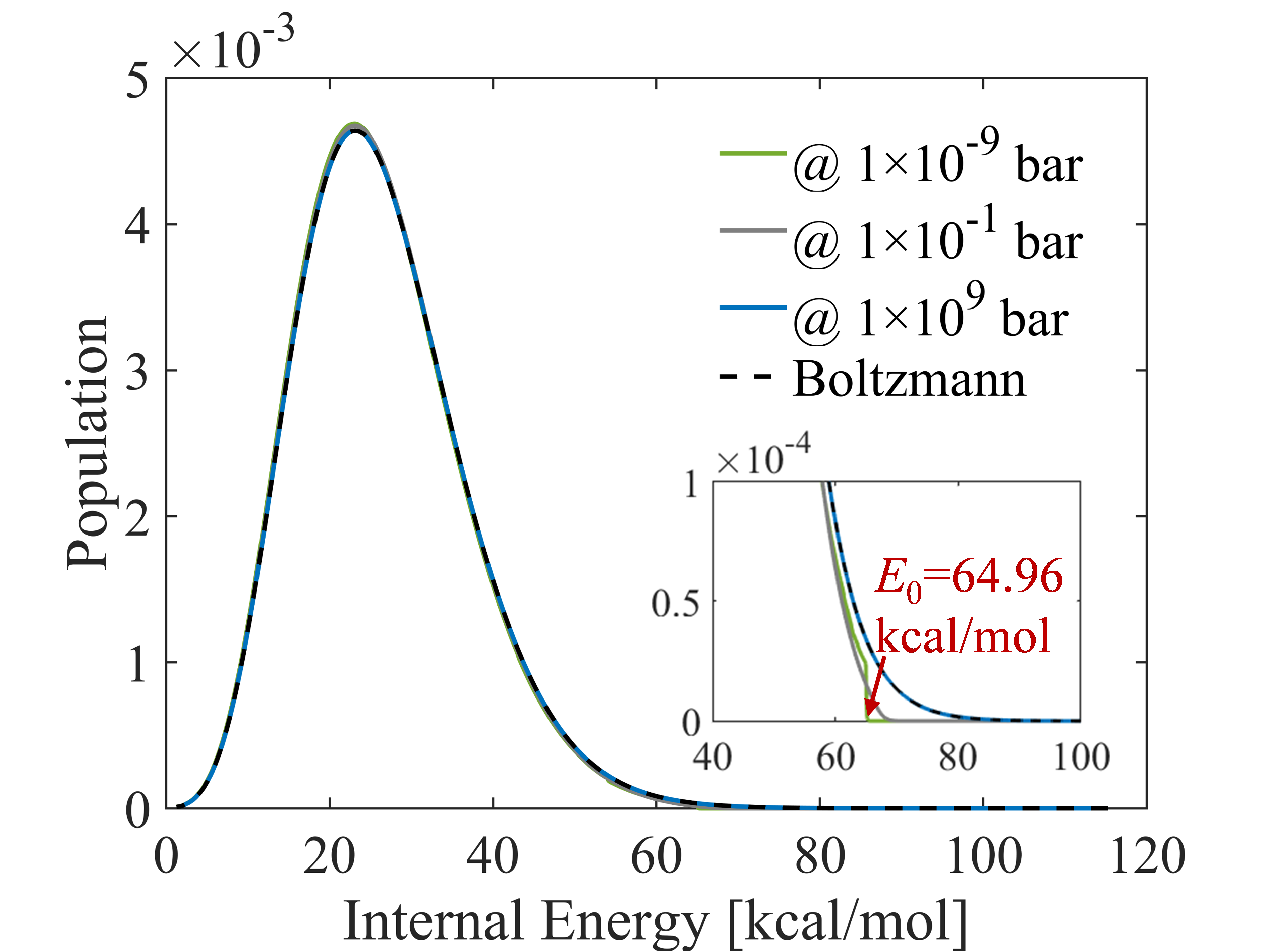}
\centering
\caption{Internal energy distribution of silicilic acid in the reaction system at three different pressures and the Boltzmann distribution at the corresponding temperature. 
}
\end{figure}

%In summary, with the ME results of R1, the collision-radiation combined activation mechanism for unimolecular kinetics is demonstrated. 
Despite its effectiveness in revealing the details of the reaction system, the ME method is computationally expensive and time-consuming. Therefore, before starting such calculations, it is desirable to evaluate whether radiation or collision activation needs to be considered. 
To this end, it is advantageous to develop a straightforward evaluating method based on the principles and framework of the ELT.

\subsection{Distinguishing radiation-dominant and collision-dominant regimes}

As the ELT method reliably describes the kinetics in regimes where radiation activation plays a competitive or dominant role, we can use it to find a more simple way to evaluate the competition between collision-induced and radiation-induced activation mechanisms under different reaction conditions. Equation \ref{eq:LindemannELT} indicates that the transition from the radiation-dominant regime to the collision-dominant regime occurs where $ k^{\rm c}_{1}\lbrack {\rm M} \rbrack = k^{\rm r}_{1}\lbrack h\nu \rbrack $. 
%It's hard to correlate the two terms directly with the pressure and temperature of the environment and the properties of the molecules. 
As shown by Eq.~\eqref{eq:kpump_collision}, the collisional contribution to the rate constant $ k^{\rm c}_{1}\lbrack {\rm M} \rbrack $ is directly proportional to the collision frequency $Z$, while the detailed analysis in Appendix A shows that the radiational contribution to the rate constant $k^{\rm r}_{1}\lbrack h\nu \rbrack $ is proportional to the total photon absorption rate $\Omega$ of the molecule that is a sum of radiation absorption over the radiation-induced transition rates of all vibrational modes.
Thus, a dimensionless number $PR$ to identify the relative importance of collision-activation and radiation activation mechanisms can be defined as
\begin{equation}\label{eq:PR}
    PR \equiv \frac{Z}{\Omega} \sim \frac{k^{\rm c}_{1}\lbrack {\rm M} \rbrack}{k^{\rm r}_{1}\lbrack h\nu \rbrack}
\end{equation}
% The \(PR\) number is calculated as the ratio of collisional frequency \(Z\) to the total radiation absorption rate \(\Omega\). 
%With the physical meaning of the proposed \(PR\) number, we can know that $PR = \frac{Z}{\Omega} \sim \frac{k^{\rm c}_{1}\lbrack {\rm M} \rbrack}{k^{\rm r}_{1}\lbrack h\nu \rbrack}$. 
%When \(PR \gg 1\) , namely, the collisional frequency is much larger than the radiation absorption rate, the reaction is dominated by the collision activation, and vice versa, by the radiation activation. 

Considering that molecules typically have IR-active vibrational modes of $\sim$10\textsuperscript{3} cm\textsuperscript{-1} (e.g. the stretch mode of the CH bond usually has a frequency of $\sim$3000 cm\textsuperscript{-1}), but the averaged collisional energy transfer is usually around $\sim$10\textsuperscript{2} cm\textsuperscript{-1}, namely, a single radiation absorption has a higher energy-transfer efficiency than a single collision, we expect that $PR^* \approx 10$ marks the boundary between the radiation-dominant and collision-dominant regimes at any specific temperature.  

Figure 5 confirms this expectation, in which we plot the $Z$/$\Omega$ values at the regime boundaries over a wide temperature range, for a number of unimolecular reactions typical in the interstellar environment \citep{Vuitton2012,Vazart2015a,Vazart2015,Puzzarini2015,Zhang2020_Cya}, including reaction R1, the unimolecular dissociation reactions of four hydrocarbons (ethane, butane, benzene, and hexane), and the unimolecular isomerization reactions between two cyanomethanimine isomers (Z-isomer and E-isomer). 
% \HXCOMM{I suggest we call the $Z/\Omega$ values shown in Figure 5 $PR^*$.}\ZXRCOMM{OK}
%These reactions can be seen as representative reactions in interstellar \citep{Vuitton2012,Vazart2015a,Vazart2015,Puzzarini2015,Zhang2020_Cya}.  
% (Corresponding computational details of these tested reactions can be found in the supporting information.)

The boundary values of $Z$/$\Omega$ shown in Figure 5 are determined as
\begin{equation}\label{eq:PR_cc}
    PR^* \equiv \frac{Z}{\Omega}\bigg\vert_{ k^{\rm c}_{1}[{\rm M}] = k^{\rm r}_{1}[h\nu]}= {\frac{Z}{\Omega}} \bigg/ {\frac{k^{\rm c}_{1}\lbrack {\rm M} \rbrack}{k^{\rm r}_{1}\lbrack h\nu \rbrack}}
\end{equation}
because the boundary is defined where $k^{\rm c}_{1}\lbrack {\rm M} \rbrack = k^{\rm r}_{1}\lbrack h\nu \rbrack$, 
and the hard-sphere collision model is used for calculating $Z$, and Eqs.~\eqref{eq:Omega1}, \eqref{eq:kpump_collision}, and \eqref{eq:kpump_radiation} are used to calculate $\Omega$, $k^{\rm c}_{1}[{\rm M}]$, and $k^{\rm r}_{1}\lbrack {\rm h\nu} \rbrack$.
We note that since the collision frequency $Z$ is proportional to the molecule number density $[{\rm M}]$, the boundary values of $Z$/$\Omega$ are pressure independent.

\begin{figure}\label{fig:PR}
\centering
%\subfigure[ The radiation absorption rate \(\Omega\) of different molecules. The X-axis represents the number of degrees of freedom (d.o.f). ]{
%\includegraphics[width=0.9\linewidth]%{Figure/FigureS1_Omega.png}
%}
%\subfigure[ The $PR$ numbers of a series of unimolecular reactions at the turning points of dominant activation mechanisms as a function of temperature]
{
\includegraphics[width=0.9\linewidth]{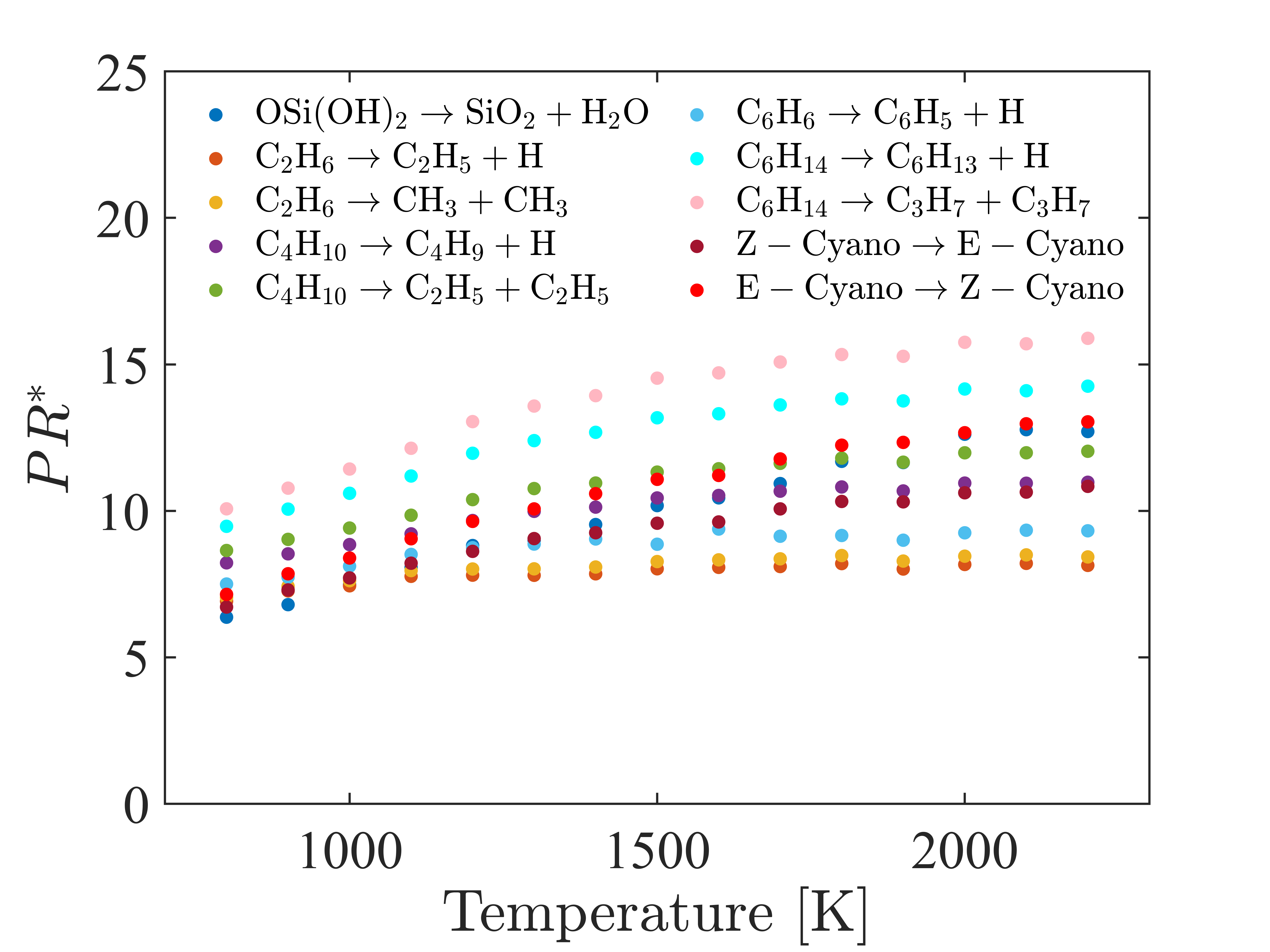}
}
\caption{
  The boundary values of $Z/\Omega$ ($PR^*$) as a function of temperature for some typical interstellar unimolecular reactions.} 
  %$PR$ numbers of a series of unimolecular reactions at the turning points of dominant activation mechanisms as a function of temperature.  }
\end{figure}

We can see that almost all the tested unimolecular reactions have a $Z$/$\Omega$ boundary value ($PR^*$  number) between 5 and 15, and these $PR^*$ numbers are weakly dependent on the temperature, the reaction type, and the size of unimolecular reactant (the changing is less than a factor of three). Thus, we propose to approximately use $PR^* = 10$ as the reference value distinguishing the collision-dominated and radiation-dominated regimes. The red line in figure 3 shows that $PR = 10$ successfully captures the starting pressure of the radiation-dominated region observed by both ME and ELT methods.

By means of the $PR$ number, we can understand why only the collision activation mechanism is commonly considered in atmospheric chemistry and combustion chemistry, but the radiation activation mechanism could play a more important role in the interstellar medium (ISM). At room temperature and atmospheric pressure, the rate of radiation absorption is usually on the order of $\unit[10^{3}]{s^{-1}}$, and the collision frequency is on the order of $\unit[10^{9}]{s^{-1}}$, thus $PR \sim 10^6 \gg PR^*$, indicating that it belongs to the collision-dominated condition and the radiation activation can be neglected.
%In this case, the equation \ref{eq:LindemannOrigin} can be approximated as the conventional Lindemann theory equation \ref{eq:LindemannP}, and this is the reason why only pressure dependence is . 
In the ISM, the pressure is fairly low, for example, the collisional frequency is about $\unit[10^{-4}]{s^{-1}}$ for a pressure around $\unit[10^{-13}]{bar}$,  thus $ PR \sim 10^{-7} \ll PR^* $, and the kinetics of unimolecular reaction is dominated by radiation in this case. 
%Here, the equation \ref{eq:LindemannOrigin} can be approximated as the equation \ref{eq:Lindemannhv}. 

In many astrophysical environments, like the surroundings of stars or the planetary atmosphere, the pressure varies over a wide range, thus collision and radiation may compete with each other in determining the kinetics, resulting in interesting change of the dominating mechanism in space.
% Thus understanding the dominance of these two effects is crucial. 
In the following subsections, we will present two such examples.

\subsection{Silicilic acid dissociation in AGB star surroundings}
As a key reaction in the AGB star surroundings, the R1 reaction kinetics is investigated along the radial position of AGB star. The calculated reaction rate constants of R1 are shown in the left panel of Figure 6 as a function of the star radial position,
including the ME results with the combined activation mechanism and the ME results with only the collision activation mechanism for comparison.

\begin{figure}[hbt!]\label{Figure:FigAGB}
\centering
\includegraphics[width=0.98\columnwidth]{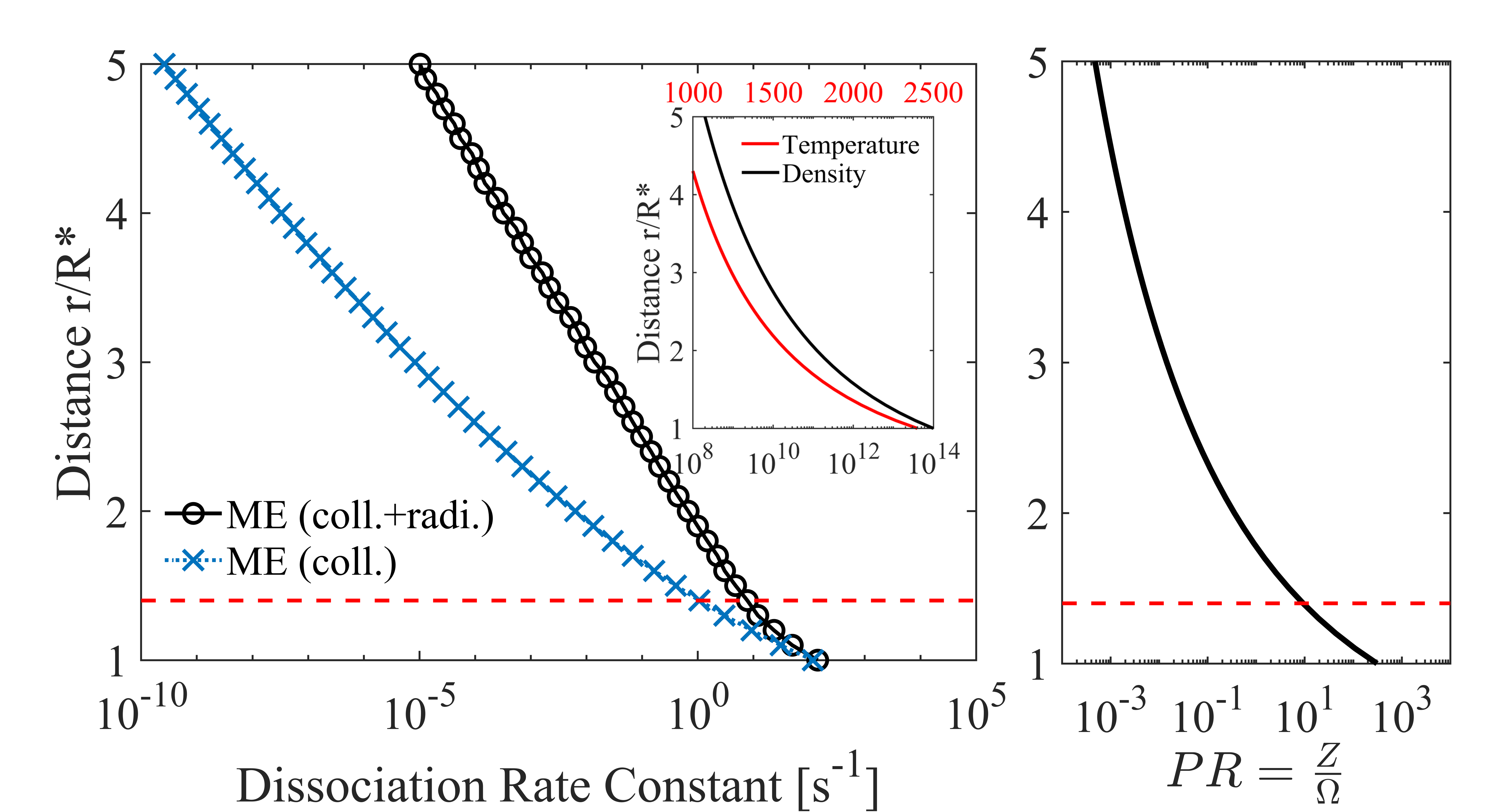}
\caption{Rate constant of silicilic acid dissociation reaction (left panel) and the local $PR$ number (right panel) as a function of radial position \textit{r} (in units of the stellar radius \textit{R}*). 
The red dash line denotes the radial position at which $PR=10$.
The insert in the left panel} shows the temperature (K) and gas density (molecule/cm\textsuperscript{3}) as a function of \textit{r} \citep{Gobrecht2016,Decin2017}.
\end{figure}

The insert in the left panel of Figure 6 shows the temperature and gas density as a function of the radial position as obtained by \citep{Gobrecht2016,Decin2017}. %\HXCOMM{Are the temperature and gas density calculated by yourself or from the cited work?} 
Around the AGB star, the temperature varies from 1000 K to 2500 K, and the gas density spans a larger range from $\unit[10^{8}]{}$ to $\unit[10^{14}]{molecule/cm^{3}}$, corresponding to a pressure of $\unit[10^{-11}]{}$ to $\unit[10^{-5}]{bar}$. 
The local $PR$ number as a function of radial position \textit{r} is shown in the right panel of Figure 6, which gives $PR=10$ at $r \approx 1.4  R^*$. It suggests that in almost all surrounding area of AGB star the R1 reaction is activated dominantly by radiation, that is, the rate constants will be underestimated considerably if only the collision activation is considered as shown in Figure 6. 
% In the radiation-dominated region, we observe that the ME in the combined mechanism and the ELT give the consistent results, again proving the good performance of the ELT method in the region where the radiation plays an important role. 

\subsection{Methyl radicals association in Titan's atmosphere} 
Titan's atmosphere is rich in a variety of hydrocarbons, and the methyl radicals association reaction (R2) as one of typical hydrocarbon reactions in Titan's atmosphere has been systematically studied \citep{Vuitton2012}. 
Vuitton et al. have found that the chemistry of Titan’s upper atmosphere is dominated by the radiation. Building on this finding, we take R2 as the second example to demonstrate the competition between radiation and collision activation. Additionally, we aim to identify and distinguish the regions where each of these mechanisms is dominant by employing our proposed method.
The rate constants of R2 and the $PR$ number are plotted as a function of altitude in Figure 7. 
%Calculation ways are the same as in section 4.1. 
In the insert of Figure 7, we also depict the change of temperature and density of the bath gas molecule N\textsubscript{2} as a function of altitude in Titan's atmosphere \citep{Krasnopolsky2009}. 
We can see that the temperature in Titan's atmosphere is very low, varying in 50 -- 200 K, and the density is around 10\textsuperscript{5} to 10\textsuperscript{20} molecule/cm\textsuperscript{3}, covering a much larger pressure range (10\textsuperscript{-14} to 10\textsuperscript{1} bar) compared to the AGB surrounding. 

\begin{figure}[hbt!]\label{Figure:FigTitan}
\centering
\includegraphics[width=0.98\columnwidth]{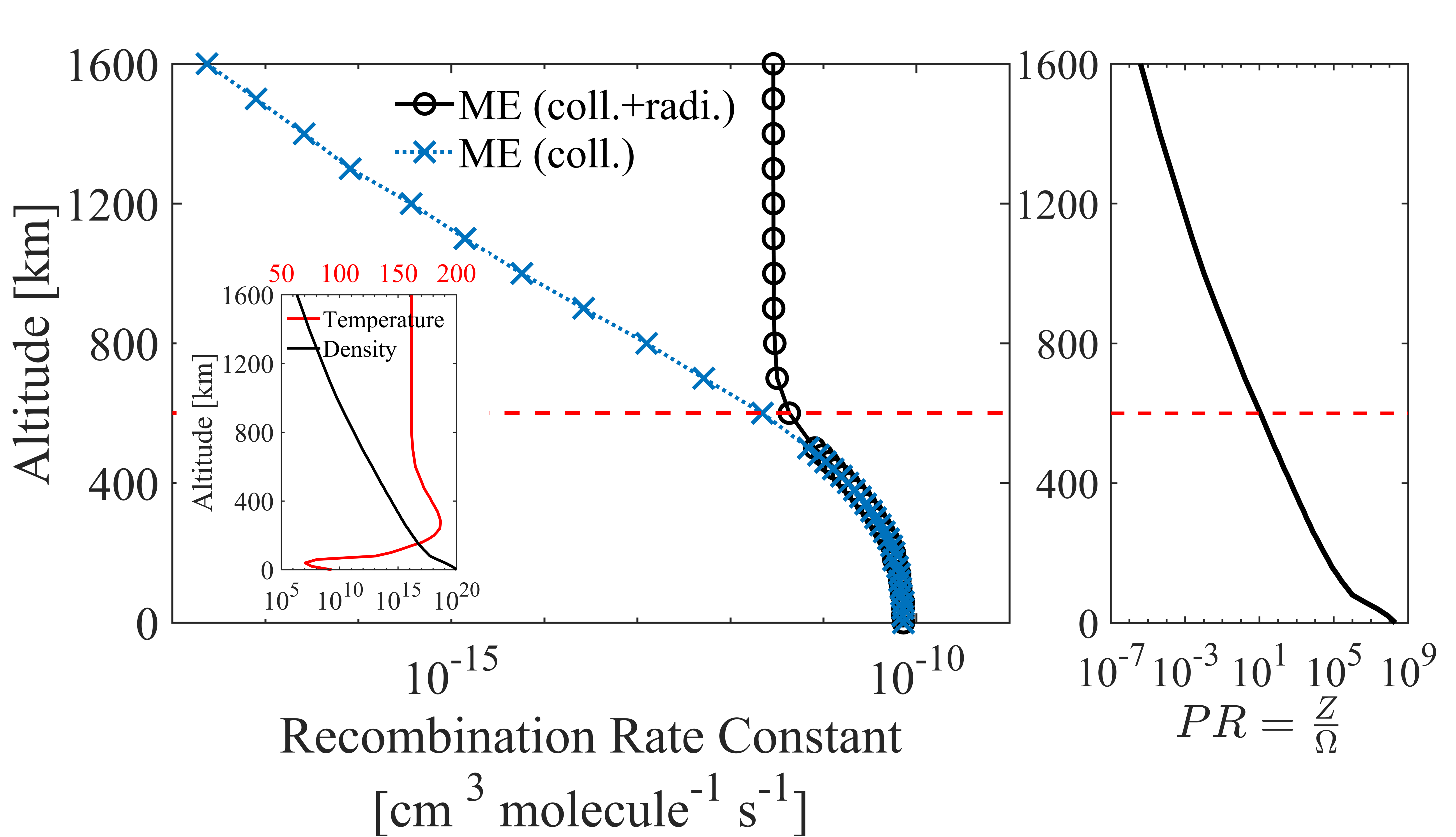}
\caption{Rate constant of methyl association reaction (R2) and the local $PR$ number in Titan's atmosphere as a function of the altitude. 
The red dash line denotes the altitude at which $PR=10$.
The insert plots the atmospheric temperature (K) and N\textsubscript{2} density (molecule/cm\textsuperscript{3}) as a function of the altitude \citep{Krasnopolsky2009}.}
\end{figure}

The red dashed line in figure 7 denotes that $PR=10$ is reached at the altitude of $\sim\unit[600]{km}$. Thus, Titan's atmosphere can be divided into collision-dominated region (below 600 km) and radiation-dominated region (above 600 km). This prediction is confirmed by the ME calculations.
Moreover, in the radiation-dominated region, at altitude above $\unit[700]{km}$, the rate constant remains constant, because the constant atmospheric temperature provides fixed environment radiation. 
In the collision-dominated region, when approaching Titan's surface, the rate constant also reaches a constant value, despite the continuous increase of pressure (gas density) towards the surface, indicating that the high-pressure limit of the collision-activation mechanism is reached.

\section{Conclusion}\label{sec:conclusion}

In interstellar conditions, radiation-activation could play a crucial role in determining the kinetics of unimolecular reactions, in contrast to the situation in terrestrial  systems, where unimolecular reactions are dominated by collision-activation because of the range of pressures. Our analysis based on the extended Lindeman theory shows that a dimensionless number $PR$, which is the ratio of the collision frequency and the IR radiation absorption rate and can be obtained with much less effort compared to full calculations using the master equations, provides a convenient and accurate measure to evaluate the relative importance of the collision-activation mechanism and the radiation-activation mechanism, as demonstrated in two example reactions in different interstellar environment. This method, i.e., using the $PR$ number to determine the dominant reaction activation mechanism, if included in the modelling of interstellar gas-phase reactions, could improve both the accuracy and cost-efficiency of the modelling.

%% Also note that the akcnowlodgment environment does not support long amounts of text. If you have a lot of people and institutions to acknowledge, do not use this command. Instead, create a new \section{Acknowledgments}.

\begin{acknowledgments}
We thank Dr. Xinyu Zheng
% , Dr. XXX and Prof. XXX 
for helpful discussions. 
We are grateful to the Natural Science Foundation of China for financial support through grants 11988102 and 21973053.
% XZ and HX are grateful to the financial support from the Natural Science Foundation of China (NSFC) through the Basic Science Center Program ``Multiscale Problems in Nonlinear Mechanics'' (grant no. 11988102). 
% XZ and XX are grateful to the financial support from the Natural Science Foundation of China (NSFC) through xxx (Award 21973053). 
\end{acknowledgments}

\appendix

\section{IR radiation induced vibrational level transition}

The IR radiation induced vibrational level transition involves three processes: spontaneous emission, stimulated emission, and absorption \citep{Plane2022,Salzburger2022} .

% \begin{figure}
% \plotone{Figure/FigureA1}
% \caption{Illustration of spontaneous emission, stimulated emission, and absorption}
% \end{figure}

The molecule may emit a photon with a frequency \(\nu\) spontaneously,
causing a vibrational energy level transition downward from a higher level \(E_{i}\) to a
lower level \(E_{j}\) meanwhile, which is called spontaneous emission,
and Einstein coefficients \(A_{ij}\) of the spontaneous emission  is used to
quantify the timescale of this process. Notice, the energy difference
between these two levels is identical to the energy of the photon.

\begin{equation}
    E_{i} - E_{j} = h\nu
\end{equation}

On the other hand, under the background radiation field, interacting
with the photon with frequency \(\nu\), the molecule may transition
downward and upward, namely from a higher vibrational level \(E_{i}\) to
a lower level \(E_{j}\), as well as from a lower vibrational level
\(E_{j}\) to a higher level \(E_{i}\). The former is called stimulated
emission, and the latter is absorption. The timescales of two processes
are \(I_{\nu}B_{ij}\) and \(I_{\nu}B_{ji}\) respectively, where
\(B_{ij}\) are \(B_{ji}\) the stimulated emission Einstein coefficients
and the absorption Einstein coefficients, and \(I_{\nu}\) is the
background radiation field intensity of the photon with frequency
\(\nu\).

Commonly, the background radiation field intensity is assumed with the
Planck distribution. So, we can obtain the radiation field intensity
\(I_{\nu}\):

\begin{equation}
I_{\nu} = \frac{8\pi h\nu^{3}}{c^{3}} \frac{1}{{\rm exp}(\frac{h\nu}{k_{\rm B}T})-1} \\
\end{equation}
where \(h\) is the Planck constant, \(c\) is the light speed.

These three Einstein coefficients are properties of molecule, and they are related by: 

\begin{equation}
A_{ij} = \frac{8\pi h\nu^{3}}{c^{3}\ }B_{ij} \quad (i>j) 
\end{equation}

\begin{equation}
B_{ij} g_{i} = B_{ji} g_{j} \quad (i>j) 
\end{equation}
where \(g_{i}\) and \(g_{j}\) are density of state of vibrational levels
\emph{i} and \emph{j} respectively.

The polyatomic nonlinear molecules may have \(3N_{\rm{atoms}} - 6\) vibrational modes,
where \(N_{\rm{atoms}}\) is the number of atoms. Under the harmonic
oscillator approximation, for a specific vibrational mode \(q\) with the
frequency \(\nu_{q}\), one can estimate the spontaneous emission
Einstein coefficient \(A_{ij,q}\) from the harmonic oscillator strength:

\begin{equation}
A_{ij,q} = \frac{8\pi c{{\widetilde{\nu}}_{q}}^{2}}{N_{A}\ }S_{q} \\
\end{equation}

where \({\widetilde{\nu}}_{q}\) is the wavenumber of the vibrational mode \emph{q} identical to \(\nu_{q}/c\), \(S_{q}\) is the harmonic oscillator strength of the vibrational mode \emph{q}, and \(N_{A}\) is Avogadro's constant. Therefore, having the harmonic oscillator strength and frequency, like from electronic structure calculation, one can obtain these three Einstein coefficients from the equations above.

We can model the radiation induced transition rates in equation \ref{eq:ME}, for downward transition:

\begin{equation}\label{eq:Rij}
R(i \rightarrow j) = \left\{ 
\begin{aligned}
A_{ij,q} + I_{\nu_{q}}B_{ij,q} \quad E_{i} - E_{j} = h\nu_{q} \\   
0 \ \ \ \ \ \ \ \ \ \ \ \ \ \ \ \  \quad \text{others} 
\end{aligned} 
\right. \quad (i>j) 
\end{equation}
and for upward transition:

\begin{equation}\label{eq:Rji}
R(j \rightarrow i) = \left\{ 
\begin{aligned}  
I_{\nu_{q}}B_{ji,q} \quad E_{i} - E_{j} = h\nu_{q} \\ 
0 \ \ \ \ \ \ \ \ \ \ \ \ \ \ \ \  \quad \text{others} 
\end{aligned} 
\right.  \quad (i>j) 
\end{equation}
where \(q = 1,2,\ldots,3N_{\rm{atoms}} - 6\).

The total radiation absorption rate \(\Omega\) is defined with these Einstein coefficients.  

\begin{equation}\label{eq:Omega1}
    \Omega = \sum_{q}^{}{ A_{ij,q} + I_{\nu_{q}}B_{ij,q} }
\end{equation}

% For downward or upward transition, they have the same magnitude commonly, so one can use either of them:
% or 
% \begin{equation}\label{eq:Omega2}
%     \Omega = \sum_{q}^{}{ I_{\nu_{q}}B_{ji,q} }
% \end{equation}

\bibliography{IRME_draft}{}
\bibliographystyle{aasjournal}

\end{document}